\documentclass[aps,prl,reprint,superscriptaddress,amsmath,amssymb]{revtex4-2}

\usepackage{docs}
\usepackage{graphicx}
\usepackage{color}

\renewcommand{\thesection}{\Roman{section}}
\renewcommand{\thesubsection}{\Alph{subsection}}

\RequirePackage{lineno}

\begin{document}


\title{Comprehensive measurement of $\eta^\prime$ photoproduction off the proton at \mbox{\boldmath $E_\gamma < 2.4$~$\mathrm{GeV}$}}



\author{N.~Muramatsu}
\affiliation{\small Institute of Modern Physics, Chinese Academy of Sciences, 
                 Lanzhou, 730000, China}
\author{J.K.~Ahn}
\affiliation{\small Department of Physics, Korea University, Seoul 02841, Republic of Korea}
\author{W.C.~Chang}
\affiliation{\small Institute of Physics, Academia Sinica, Taipei 11529, Taiwan}
\author{J.Y.~Chen}
\affiliation{\small National Synchrotron Radiation Research Center, Hsinchu 30076,
                 Taiwan}
\author{M.L.~Chu}
\affiliation{\small Institute of Physics, Academia Sinica, Taipei 11529, Taiwan}
\author{S.~Dat\'{e}}
\affiliation{\small Research Center for Nuclear Physics, Osaka University, Ibaraki,
                 Osaka 567-0047, Japan}
\author{T.~Gogami}
\affiliation{\small Department of Physics, Kyoto University, Kyoto 606-8502, Japan}
%
%
\author{H.~Hamano}
\affiliation{\small Research Center for Nuclear Physics, Osaka University, Ibaraki,
                 Osaka 567-0047, Japan}
\author{T.~Hashimoto}
\affiliation{\small Research Center for Nuclear Physics, Osaka University, Ibaraki,
                 Osaka 567-0047, Japan}
\author{Q.H.~He}
\affiliation{\small Department of Nuclear Science and Technology, College of Materials 
                 Science and Technology, Nanjing University of Aeronautics and Astronautics, 
                 Nanjing 210016, China}
\author{K.~Hicks}
\affiliation{\small Department of Physics and Astronomy, Ohio University, Athens,
                 OH 45701, USA}
\author{T.~Hiraiwa}
\affiliation{\small RIKEN SPring-8 Center, Sayo, Hyogo 679-5148, Japan}
\author{Y.~Honda}
\affiliation{\small Research Center for Accelerator and Radioisotope Science, Tohoku 
                 University, Sendai, Miyagi 982-0826, Japan}
\author{T.~Hotta}
\affiliation{\small Research Center for Nuclear Physics, Osaka University, Ibaraki,
                 Osaka 567-0047, Japan}
%
%
\author{Y.~Inoue}
\affiliation{\small Research Center for Accelerator and Radioisotope Science, Tohoku 
                 University, Sendai, Miyagi 982-0826, Japan}
\author{T.~Ishikawa}
\affiliation{\small Research Center for Nuclear Physics, Osaka University, Ibaraki,
                 Osaka 567-0047, Japan}
\author{I.~Jaegle}
\affiliation{\small Thomas Jefferson National Accelerator Facility, Newport News, Virginia 23606. USA}
%
%
\author{Y.~Kasamatsu}
\affiliation{\small Research Center for Nuclear Physics, Osaka University, Ibaraki,
                 Osaka 567-0047, Japan}
\author{V.~L.~Kashevarov}
\affiliation{\small Institut f\"{u}r Kernphysik, Johannes Gutenberg-Universit\"{a}t Mainz, D-55099 Mainz, Germany}
\author{H.~Katsuragawa}
\affiliation{\small Research Center for Nuclear Physics, Osaka University, Ibaraki,
                 Osaka 567-0047, Japan}
\author{S.~Kido}
\affiliation{\small Research Center for Accelerator and Radioisotope Science, Tohoku 
                 University, Sendai, Miyagi 982-0826, Japan}
\author{R.~Kobayakawa}
\affiliation{\small Research Center for Nuclear Physics, Osaka University, Ibaraki,
                 Osaka 567-0047, Japan}
\author{Y.~Kon}
\affiliation{\small Research Center for Nuclear Physics, Osaka University, Ibaraki,
                 Osaka 567-0047, Japan}
\author{S.~Masumoto}
\affiliation{\small Department of Physics, University of Tokyo, Tokyo 113-0033, Japan}
\author{Y.~Matsumura}
\affiliation{\small Research Center for Accelerator and Radioisotope Science, Tohoku 
                 University, Sendai, Miyagi 982-0826, Japan}
\author{M.~Miyabe}
\affiliation{\small Research Center for Accelerator and Radioisotope Science, Tohoku 
                 University, Sendai, Miyagi 982-0826, Japan}
\author{K.~Mizutani}
\affiliation{\small Research Center for Nuclear Physics, Osaka University, Ibaraki,
                 Osaka 567-0047, Japan}
\author{T.Z.~Mo}
\affiliation{\small Department of Nuclear Science and Technology, College of Materials 
                 Science and Technology, Nanjing University of Aeronautics and Astronautics, 
                 Nanjing 210016, China}
\author{T.~Nakamura}
\affiliation{\small Department of Education, Gifu University, Gifu 501-1193, Japan}
\author{T.~Nakano}
\affiliation{\small Research Center for Nuclear Physics, Osaka University, Ibaraki,
                 Osaka 567-0047, Japan}
\author{T.~Nam}
\affiliation{\small Dalat Nuclear Research Institute, Dalat, Lam Dong, Vietnam}
\author{M.~Niiyama}
\affiliation{\small Department of Physics, Kyoto Sangyo University, Kyoto 603-8555, Japan}
\author{K.~V.~Nikonov}
\affiliation{\small NRC Kurchatov Institute, Petersburg Nuclear Physics Institute, 
                 Gatchina 188300, Russia}
\author{Y.~Nozawa}
\affiliation{\small Department of Radiology, The University of Tokyo Hospital, 
                 Tokyo 113-8655, Japan}
\author{Y.~Ohashi}
\affiliation{\small Research Center for Nuclear Physics, Osaka University, Ibaraki,
                 Osaka 567-0047, Japan}
\author{H.~Ohnishi}
\affiliation{\small Research Center for Accelerator and Radioisotope Science, Tohoku 
                 University, Sendai, Miyagi 982-0826, Japan}
\author{T.~Ohta}
\affiliation{\small Department of Radiology, The University of Tokyo Hospital, 
                 Tokyo 113-8655, Japan}
\author{M.~Okabe}
\affiliation{\small Research Center for Accelerator and Radioisotope Science, Tohoku 
                 University, Sendai, Miyagi 982-0826, Japan}
\author{K.~Ozawa}
\affiliation{\small Institute of Particle and Nuclear Studies, High Energy Accelerator
                 Research Organization (KEK), Tsukuba, Ibaraki 305-0801, Japan}
\author{C.~Rangacharyulu}
\affiliation{\small Department of Physics and Engineering Physics, University of 
                 Saskatchewan, Saskatoon, Canada SK S7N 5E2}
\author{S.Y.~Ryu}
\affiliation{\small Research Center for Nuclear Physics, Osaka University, Ibaraki,
                 Osaka 567-0047, Japan}
\author{Y.~Sada}
\affiliation{\small Research Center for Accelerator and Radioisotope Science, Tohoku 
                 University, Sendai, Miyagi 982-0826, Japan}
\author{A.~V.~Sarantsev}
\affiliation{\small NRC Kurchatov Institute, Petersburg Nuclear Physics Institute, 
                 Gatchina 188300, Russia}
\author{T.~Shibukawa}
\affiliation{\small Department of Physics, University of Tokyo, Tokyo 113-0033, Japan}
\author{H.~Shimizu}
\affiliation{\small Research Center for Accelerator and Radioisotope Science, Tohoku 
                 University, Sendai, Miyagi 982-0826, Japan}
\author{R.~Shirai}
\affiliation{\small Research Center for Accelerator and Radioisotope Science, Tohoku 
                 University, Sendai, Miyagi 982-0826, Japan}
\author{K.~Shiraishi}
\affiliation{\small Research Center for Accelerator and Radioisotope Science, Tohoku 
                 University, Sendai, Miyagi 982-0826, Japan}
\author{E.A.~Strokovsky}
\affiliation{\small Laboratory of High Energy Physics, Joint Institute for Nuclear Research, 
                 Dubna, Moscow Region, 142281, Russia}
%
\author{Y.~Sugaya}
\affiliation{\small Research Center for Nuclear Physics, Osaka University, Ibaraki,
                 Osaka 567-0047, Japan}
\author{M.~Sumihama}
\affiliation{\small Department of Education, Gifu University, Gifu 501-1193, Japan}
\affiliation{\small Research Center for Nuclear Physics, Osaka University, Ibaraki,
                 Osaka 567-0047, Japan}
\author{S.~Suzuki}
\affiliation{\small Japan Synchrotron Radiation Research Institute (SPring-8), Sayo,
                 Hyogo 679-5198, Japan}
\affiliation{\small Research Center for Nuclear Physics, Osaka University, Ibaraki,
                 Osaka 567-0047, Japan}
\author{S.~Tanaka}
\affiliation{\small Research Center for Nuclear Physics, Osaka University, Ibaraki,
                 Osaka 567-0047, Japan}
\author{Y.~Taniguchi}
\affiliation{\small Research Center for Accelerator and Radioisotope Science, Tohoku 
                 University, Sendai, Miyagi 982-0826, Japan}
\author{A.~Tokiyasu}
\affiliation{\small Research Center for Accelerator and Radioisotope Science, Tohoku 
                 University, Sendai, Miyagi 982-0826, Japan}
\author{N.~Tomida}
\affiliation{\small Department of Physics, Kyoto University, Kyoto 606-8502, Japan}
%
\author{Y.~Tsuchikawa}
\affiliation{\small J-PARC Center, Japan Atomic Energy Agency, Tokai, Ibaraki 319-1195, Japan}
\author{T.~Ueda}
\affiliation{\small Research Center for Accelerator and Radioisotope Science, Tohoku 
                 University, Sendai, Miyagi 982-0826, Japan}
\author{T.F.~Wang}
\affiliation{\small School of Physics, Beihang University, Beijing 100191, China}
\author{H.~Yamazaki}
\affiliation{\small Radiation Science Center, High Energy Accelerator Research Organization 
                  (KEK), Tokai, Ibaraki 319-1195, Japan}
\author{R.~Yamazaki}
\affiliation{\small Research Center for Accelerator and Radioisotope Science, Tohoku 
                 University, Sendai, Miyagi 982-0826, Japan}
\author{Y.~Yanai}
\affiliation{\small Research Center for Nuclear Physics, Osaka University, Ibaraki,
                 Osaka 567-0047, Japan}
\author{T.~Yorita}
\affiliation{\small Research Center for Nuclear Physics, Osaka University, Ibaraki,
                 Osaka 567-0047, Japan}
\author{C.~Yoshida}
\affiliation{\small Instrumentation Technology Development Center, High Energy Accelerator
                 Research Organization (KEK), Tsukuba, Ibaraki 305-0801, Japan}
\author{M.~Yosoi}
\affiliation{\small Research Center for Nuclear Physics, Osaka University, Ibaraki,
                 Osaka 567-0047, Japan}
\collaboration{BGOegg collaboration}




\begin{abstract}
For the spectroscopy of nucleon resonances at the total energies from the $\eta^\prime$-meson production threshold to $2.32$~$\mathrm{GeV}$, photon beam asymmetries of the reaction $\gamma p \to \eta^\prime p$ were measured together with total and differential cross sections by analyzing the two decay modes $\eta^\prime \to \gamma \gamma$ and $\pi^0 \pi^0 \eta$. New constraints for amplitude decomposition were given by the first-time result of photon beam asymmetries at $E_\gamma > 1.84$~$\mathrm{GeV}$ and the most precise data of differential cross sections to date at extremely backward $\eta^\prime$ angles. The possibility of a larger coupling constant of the $\eta^\prime$-nucleon system to the $N(2250)$ resonance was implied in the partial wave analyses using the present data.
\end{abstract}


\maketitle
\clearpage


{\it Introduction}--Baryon spectroscopy is important to understand the non-perturbative QCD at low energies and the quark confinement inside a hadron. Further accumulation of the experimental data for the mass range around $2$~$\mathrm{GeV}$ is now desired because the baryon spectra obtained from the existing data have not been well explained by constituent quark models and lattice QCD calculations \cite{baryspec}. This situation may reflect non-trivial hadron structures beyond the picture of baryons made of three uncorrelated constituent quarks. Here, photoproduction of a heavy pseudoscalar meson is a valuable tool to investigate high-mass baryon resonances contributing in the $s$-channel: firstly, pseudoscalar meson photoproduction is simply described by four complex amplitudes (e.g., CGLN and helicity amplitudes) \cite{piphoto}; secondly, photon beam polarization is effective to solve those amplitudes and decompose overlapping resonances by measuring polarization observables, which differ from the cross section in the amplitude-based representation \cite{piphoto, poldata}; thirdly, amplitude analyses for highly excited baryons become easier using the two-body final state of a heavy meson and a proton instead of multi-meson photoproduction.

This letter focuses on the reaction $\gamma p \to \eta^\prime p$, where the $\eta^\prime$ meson has an extraordinarily large mass among ground-state pseudoscalar mesons because of the $U_A$(1) quantum anomaly. Since its isospin is zero, only nucleon resonances ($N^*$) appear in the $s$-channel. It is also possible to explore the $N^*$s’ coupling with $s \bar{s}$ quarks contained in the $\eta^\prime$ meson. In spite of these unique features, the experimental data of $\eta^\prime$ photoproduction are still scarce even with the proton target. So far, differential cross sections ($d\sigma/d\Omega$) have been measured by the CLAS \cite{dcsclas} and CBELSA/TAPS \cite{dcselsa} collaborations up to the total energies ($W$) of $2.840$ and $2.360$~$\mathrm{GeV}$, respectively, but their values show systematic differences in magnitude, motivating new measurements. As for the total cross section ($\sigma_{tot}$), the data exist only from the CBELSA/TAPS collaboration \cite{dcselsa}. The results of polarization observables are particularly limited; photon beam asymmetries ($\Sigma$), which represent the azimuthal angle asymmetry of the reaction plane relative to the polarization vector of a linearly polarized photon beam, are available only near the production threshold ($1.896 < W < 1.917$~$\mathrm{GeV}$) or below $W = 2.080$~$\mathrm{GeV}$ from the Graal \cite{pbagraal} and CLAS \cite{pbaclas} collaborations, respectively. Unfortunately, such limitation of the data availability for polarization observables hinders the progress of baryon spectroscopy because they are essential to differentiate the amplitude models that cannot be distinguished solely by $d\sigma/d\Omega$'s, as suggested in theoretical calculations \cite{mesex}. Thus, the measurement of $\Sigma$'s was performed by the present analysis in the unexplored energy region up to $W = 2.316$~$\mathrm{GeV}$ ($E_\gamma = 2.389$~$\mathrm{GeV}$). In addition, $d\sigma/d\Omega$'s at extremely backward $\eta^\prime$ angles were obtained with the highest precision to date.

{\it Experiment}--The experiment was carried out by the BGOegg collaboration at the SPring-8 LEPS2 beamline, where a tagged photon beam with high linear polarization was available in the energy range of $1.26$--$2.39$~$\mathrm{GeV}$ via laser Compton scattering \cite{leps2bl}. The linear polarization is maximized to $92$\% at the highest beam energy, while it drops to $51$\% at the $\eta^\prime$ production threshold. The beam energy was measured event by event with the momentum analysis of a recoil electron in Compton scattering using a tagging counter. A plastic scintillator placed upstream of main detectors worked as a veto counter for $e^+e^-$ contamination in the photon beam.

Details of the experimental setup can be found in Refs.~\cite{pi0egg, etaegg, eggtest}. The photon beam was irradiated onto a $54$~$\mathrm{mm}$-thick liquid hydrogen target, and particles produced in the $\gamma p$ reaction were measured by the detectors surrounding this target. The polar angles of $24^\circ$--$144^\circ$ were covered by the BGOegg calorimeter, where $1320$ bismuth germanate (BGO) crystals were arranged in an egg-shape to get the energies and directions of $\gamma$'s originating from meson decays. Note that its azimuthal coverage has rotational symmetry around the beam axis. The mass resolutions of reconstructed mesons were achieved to be the best among the electromagnetic calorimeters used for hadron experiments in similar energy ranges \cite{had2023}. Charged particles were also detected in the BGOegg calorimeter together with the requirement of on-time hits at the geometrically matching positions of inner plastic scintillators (IPS). A few \% of $\gamma$'s were inevitably misidentified as charged particles by IPS because of $e^+e^-$ conversions at inner materials. The forward acceptance at the polar angles less than $22^\circ$ was covered by a planar drift chamber (DC) to detect charged particles. DC has a hexagonal sensitive area, where six planes of sense wires are stretched in three directions with a relative angle of $60^\circ$. A straight track corresponding to a charged particle was reconstructed at DC.

{\it Analysis}--The data collected in 2014 were analyzed using the methods adopted in the published articles \cite{pi0egg, omgegg, etaegg, f0egg}. The integrated luminosity of the analyzed data reaches $501$~$\mathrm{nb^{-1}}$ for the energy range above the $\eta^\prime$ production threshold. Signals of the reaction $\gamma p \to \eta^\prime p$ were identified with the subsequent $\eta^\prime$ decay into $\gamma \gamma$ or $\pi^0 \pi^0 \eta \to \gamma \gamma \gamma \gamma \gamma \gamma$. The final-state $\gamma$ producing an electromagnetic shower in the BGOegg calorimeter was reconstructed as a neutral cluster by connecting neighboring crystals with energy deposit and confirming no association with an IPS hit. Then, events with two or six neutral clusters were selected after requiring the timing consistency of them (less than $10$~$\mathrm{ns}$) and omitting the clusters whose energy was below the predetermined thresholds ($30$ and $20$~$\mathrm{MeV}$ for the $2\gamma$ and $6\gamma$ modes, respectively). In addition, the detection of only one charged particle was required at DC or the BGOegg calorimeter. In the present analysis, the following special treatments were adopted to increase the signal statistics: (1) a neutral cluster whose core crystal with a maximum energy was found at the forward edge of the BGOegg calorimeter was accepted as a $\gamma$ hit with a large leak correction; (2) for the $6\gamma$ mode, the cluster-charge identification using IPS was skipped if a charged particle was detected at DC; (3) DC tracks reconstructed near the beam axis were not counted as additional particles to avoid the over-veto caused by off-timing $e^+e^-$ conversions at the target.

Selected events were subject to a 4 or 7-constraint kinematic fit in the $2\gamma$ and $6\gamma$ modes, respectively. In both modes, the fit required the four-momentum conservation between the initial and final states of the reaction $\gamma p \to \gamma \gamma p$ or $\gamma p \to \gamma \gamma \gamma \gamma \gamma \gamma p$. In the case of the $6 \gamma$ mode, three combinations of $\gamma \gamma$ were constrained to have the nominal mass of $\pi^0$ or $\eta$ \cite{pdg} while repeating fits in 45 ways of assignment. A set giving the best $\chi^2$ was adopted for the further analysis. The fits were performed by treating the magnitude of a charged-particle momentum as an unmeasured variable with the assumption of a proton because only its direction was measured. Finally, the $\chi^2$ probability of the kinematic fit was required to be greater than $2$\% and $5$\% for the $2\gamma$ and $6\gamma$ modes, respectively. 

\begin{figure}[tp]
 \centering
 \includegraphics[width=8cm]{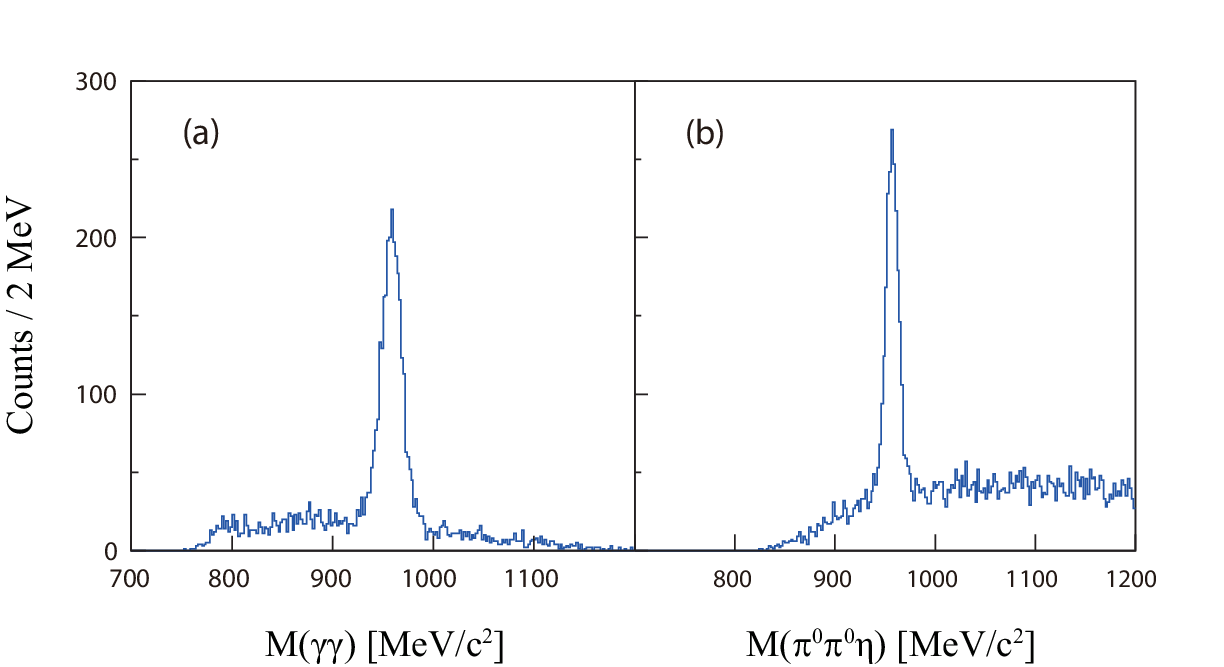}
 \caption{Invariant mass distributions of (a) $\gamma \gamma$ and (b) $\pi^0 \pi^0 \eta$ in the samples selected for the $2\gamma$ and $6\gamma$ modes, respectively.}
 \label{fig:invm}
\end{figure}
In the selected event samples, $\gamma \gamma$ and $\pi^0 \pi^0 \eta$ invariant mass distributions were obtained as shown in Fig.~\ref{fig:invm}, using the final-state particle's four-momenta improved by the kinematic fit. The $\eta^\prime$ mass resolutions in the $2\gamma$ and $6\gamma$ modes were excellent, reaching $10.5$ and $6.3$~$\mathrm{MeV/c^2}$, respectively. A broad background distribution under the $\eta^\prime$ peak remains in both modes, so that such contribution must be subtracted to count signal events in the invariant mass distribution. In the $2\gamma$ mode, most of the background comes from the reaction $\gamma p \to \pi^0 \pi^0 p \to \gamma \gamma \gamma \gamma p$ with two $\gamma$'s escaping from the calorimeter acceptance. The background shape was estimated by adding appropriately normalized $\gamma \gamma$ invariant mass spectra for the processes of $\pi^0 \Delta^+$, $\pi^0 N(1520, 1535)^+$, $\pi^0 N(1650, 1675, 1680)^+$, and non-resonant $\pi^0 \pi^0 p$ production, which were generated using Geant4 \cite{geant4} Monte Carlo (MC) simulations with the implementation of proper detector geometries and responses \cite{pi0egg, omgegg}. Here, undistinguishable $\pi^0 p$ resonances from $N(1520)^+$ and $N(1535)^+$ as well as $N(1650)^+$, $N(1675)^+$, and $N(1680)^+$ were treated as a summed resonance peak, whose mean and width were determined by the fit using the real data sample. For the $6\gamma$ mode, non-resonant $\pi^0 \pi^0 \eta$ photoproduction was considered as a main background, and its invariant mass shape was estimated from the MC simulation. The $\eta^\prime$ peak shapes were also simulated for both modes, and the signal counts in the selected event samples were extracted by fitting the simulated template spectra of the signal and background processes simultaneously. Finally, total signal counts for the $2\gamma$ and $6\gamma$ modes amount to $2.5\mathrm{K}$ and $1.9\mathrm{K}$ events, respectively.

{\it Differential cross section}--The $d\sigma/d\Omega$'s were measured in 7 energy and 10 polar angle bins for $1.896 < W < 2.316$~$\mathrm{GeV}$ and $-1 < \cos \theta^{c.m.}_{\eta^\prime} < 1$, respectively. The opening angle of a $\gamma \gamma$ pair from the $\eta^\prime$ decay tends to be large compared with the forward acceptance hole of the BGOegg calorimeter, so that smaller $\eta^\prime$ polar angles can be covered in the $2\gamma$ mode. Signal counts evaluated from the template fits in the individual kinematic bins were corrected by the geometrical acceptance factors and efficiencies obtained by the MC simulation of $\eta^\prime$ photoproduction. Additional corrections for the proton and $\gamma$ detection efficiencies were further applied to fix the differences between the real and MC data estimations \cite{pi0egg, omgegg}. The corrected signal amounts were then divided by luminosities, calculated based on the target length and the number of beam photons in each energy bin. The number of photons reaching the target was evaluated as follows: (1) the scaler count of tagging counter hits was integrated with the correction for dead time to obtain the total number of tagged photons, (2) this total number was then distributed to the individual energy bin based on the fraction in the simulated $E_\gamma$ spectrum, and (3) the photon count distributed in each energy bin was further multiplied by the energy-dependent photon-beam transmittance due to the beamline structure over the distance of $125~\mathrm{m}$ \cite{pi0egg, etaegg}. Finally, $d\sigma/d\Omega$'s were obtained by taking into account the solid angle of each polar-angle bin ($2\pi \times 0.2$) and the branching fractions of $\eta^\prime \to \gamma \gamma $ and $\eta^\prime \to \pi^0 \pi^0 \eta \to \gamma \gamma \gamma \gamma \gamma \gamma$ decays \cite{pdg}.

In the present analysis, two $\eta^\prime$ decay modes were separately treated at the first step to evaluate $d\sigma/d\Omega$’s. For backward $\eta^\prime$ angles, the statistical uncertainties of $d\sigma/d\Omega$'s in the $2\gamma$ and $6\gamma$ modes are similar to each other because their signal counts proportional to the product of the branching fraction and the geometrical acceptance are balanced. The $d\sigma/d\Omega$'s obtained in the two modes were statistically consistent with each other, so they were combined by taking a weighted mean of acceptance-corrected signal counts based on the weights made by the inversed squares of statistical uncertainties and dividing this mean by the luminosity in each kinematic bin. The statistical uncertainty for the weighted mean was evaluated by taking a square root for the inverse of the weight sum.

The systematic uncertainties of $d\sigma/d\Omega$'s in the individual kinematic bins were obtained for the sources arising from the evaluation procedure of the photon beam transmittance ($2.8$--$3.2$\%), the accuracy of the target length ($1.3$\%), the ambiguity of a photon beam position on the target ($2.0$--$3.3$\%), and the uncertainties of the branching fractions ($0.7$--$2.3$\%). (See some more details in \cite{pi0egg}.) In addition, possible variation of the signal counts depending on the choice of analysis methods was estimated by changing the $\chi^2$ probability cut position to $10$\% ($1.8$--$3.2$\%), altering the simulated signal template shapes to a Voigt function with free mean and sigma ($0.7$--$3.0$\%), and replacing the simulated background template shapes to a second-order polynomial function with free parameters ($0.4$--$3.4$\%). In these estimations, the statistics of examination samples were appropriately increased by merging neighboring kinematic bins. The uncertainties independent for the $2\gamma$ and $6\gamma$ modes were combined by taking their weighted mean based on the statistical uncertainties of $d\sigma/d\Omega$'s. Finally, all the systematic uncertainties were quadratically summed, resulting in the total uncertainties of $5.1$--$6.8$\%. The estimated systematic uncertainties were shown by hatched histograms in the following figures related to cross sections.

\begin{figure*}[tp]
 \centering
 \includegraphics[width=15cm]{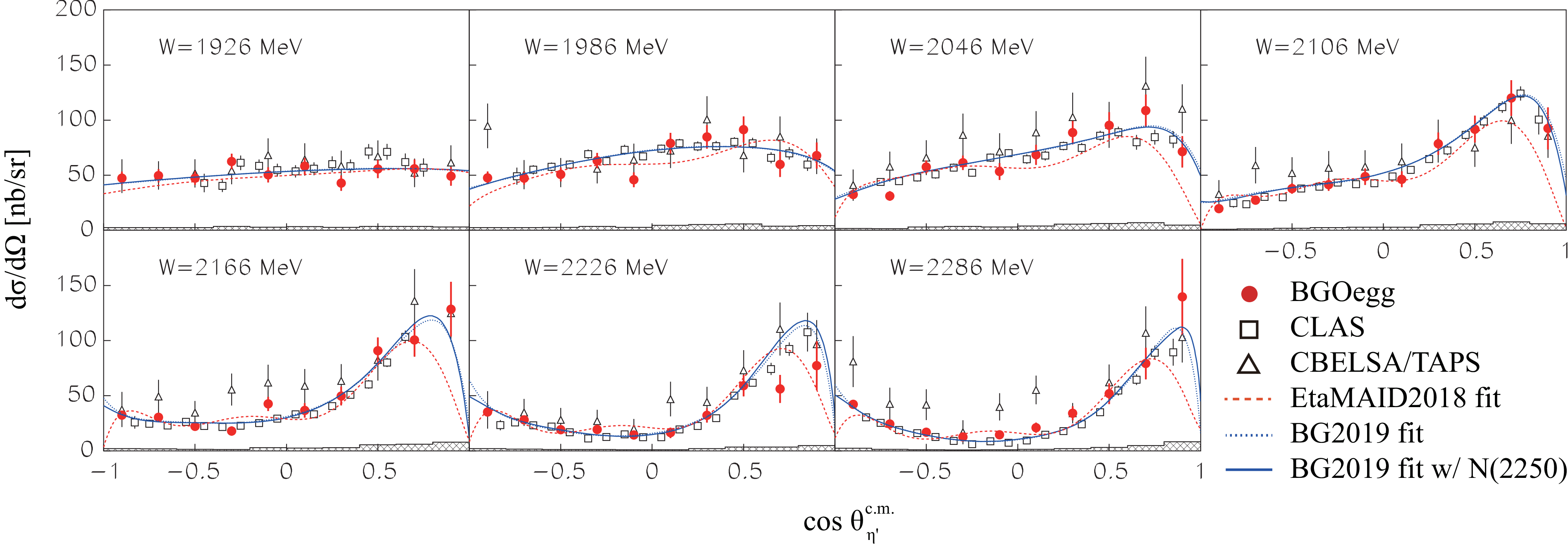}
 \caption{Differential cross sections $d\sigma/d\Omega$ of the reaction $\gamma p \to \eta^\prime p$ depending on the cosine of the $\eta^\prime$ polar angle in the center-of-mass frame, $\cos \theta_{\eta^\prime}^{c.m.}$. Data points with statistical uncertainties (vertical bars) for the present analysis (BGOegg) and the existing results are plotted by the symbols in the legend. Systematic uncertainties for the present result are indicated by the hatched histograms. Two partial wave analysis fits using the present result are also overlaid as explained in the last part of the text.}
 \label{fig:dcs}
\end{figure*}
\begin{figure*}[tp]
 \centering
 \includegraphics[width=15cm]{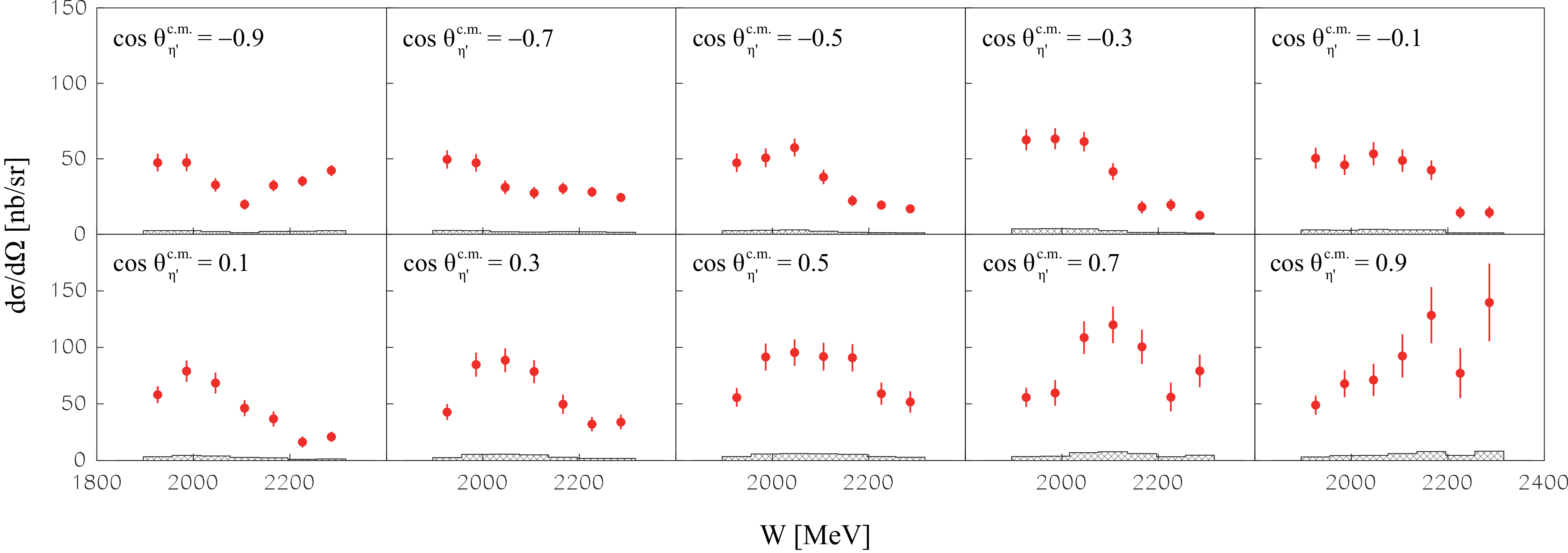}
 \caption{Differential cross sections $d\sigma/d\Omega$ of the reaction $\gamma p \to \eta^\prime p$ as a function of the total energy $W$. Only the present result in Fig.~\ref{fig:dcs} is rearranged with the same notations.}
 \label{fig:edep}
\end{figure*}

Figure~\ref{fig:dcs} shows the $d\sigma/d\Omega$'s measured by the present analysis (red closed circles with thick vertical bars, representing statistical uncertainties) and other experiments indicated in the legend. Forward enhancement coming from $t$-channel diagrams is clearly seen at higher energies, while the polar angle dependence is not strong near the threshold. The present result is statistically consistent with the CLAS values \cite{dcsclas} in the overlapped polar angle range of $\cos \theta^{c.m.}_{\eta^\prime} > -0.8$. The CBELSA/TAPS measurement \cite{dcselsa} shows systematically higher values although their statistical uncertainties are relatively large. In Fig.~\ref{fig:edep}, the present result of $d\sigma/d\Omega$'s is rearranged as a function of the total energy $W$. While the differential cross sections get lower at higher energies in $-0.8 < \cos \theta^{c.m.}_{\eta^\prime} \lesssim 0.4$, a higher energy enhancement distinctively appears at the most backward bin ($\cos \theta^{c.m.}_{\eta^\prime} = -0.9$), where the present result provides the highest precision. Such behavior is also seen in the $\eta$ photoproduction measurement by the BGOegg collaboration \cite{etaegg}, possibly implying the existence of high-spin resonances or some interference effect as discussed there.

\begin{figure}[tp]
 \centering
 \includegraphics[width=6cm]{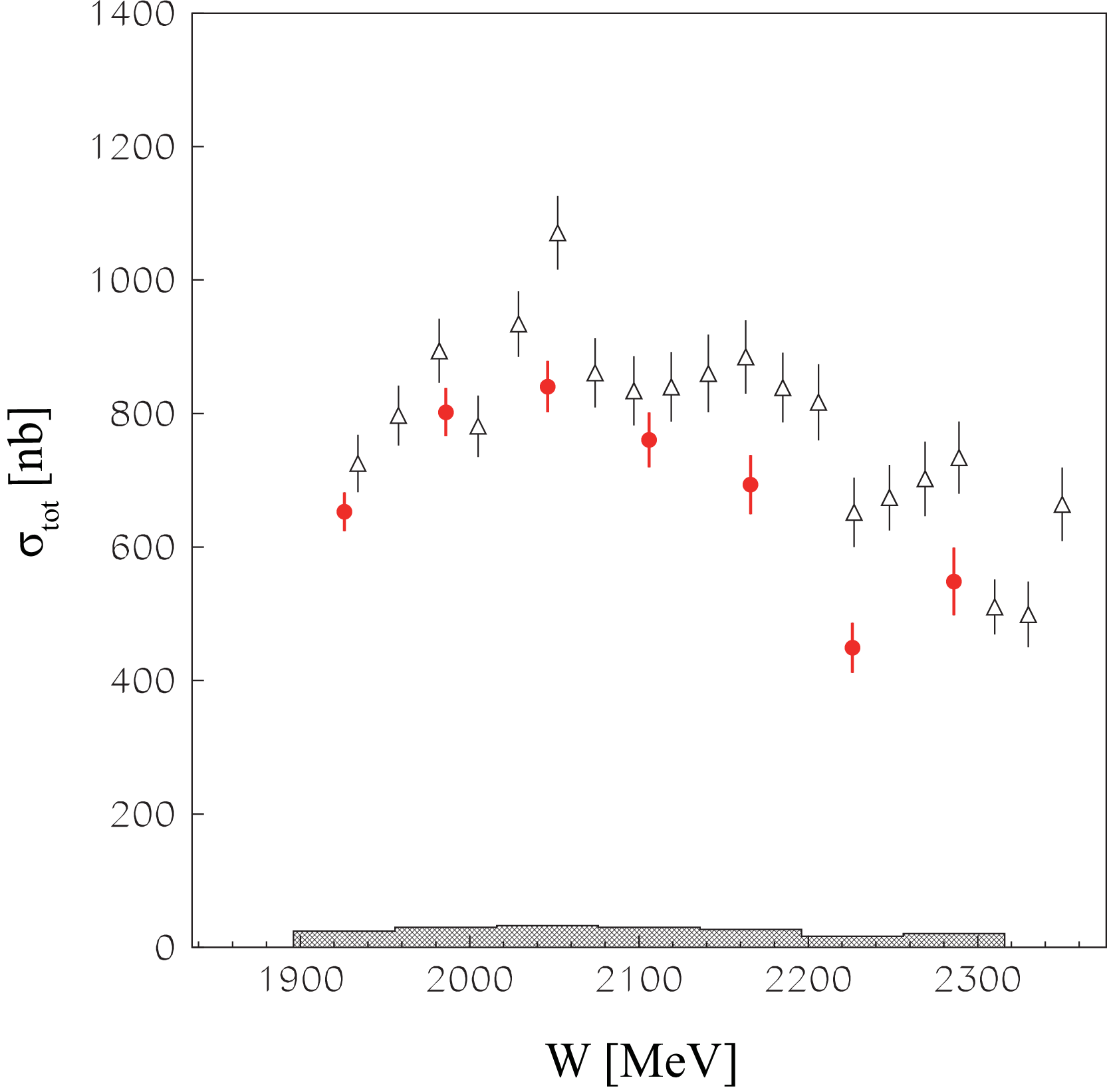}
 \caption{Total cross sections $\sigma_{tot}$ of the reaction $\gamma p \to \eta^\prime p$, measured by the present analysis (red closed circles) and the CBELSA/TAPS experiment (black open triangles). The notations are the same as those in Fig.~\ref{fig:dcs}.}
 \label{fig:tcrs}
\end{figure}

{\it Total cross section}--The $d\sigma/d\Omega$'s were measured for all the angular range, so the $\sigma_{tot}$ was simply obtained by adding the products of the $d\sigma/d\Omega$ and the corresponding solid angle. The statistical and systematic uncertainties were calculated by taking a quadratic sum of the components independently measured in the individual polar angle bins and further taking into account the common uncertainties related to the luminosity. Red closed circles with thick vertical bars in Fig.~\ref{fig:tcrs} show the $\sigma_{tot}$'s and their statistical uncertainties, respectively, measured by the present analysis. In addition, the preceding result from the CBELSA/TAPS collaboration is overlaid by open triangles \cite{dcselsa}. Both data indicate that the $\sigma_{tot}$ reaches a maximum around $2040$~$\mathrm{MeV}$ and gradually decreases at higher energies. However, the two results are systematically inconsistent, reflecting the deviation observed in the $d\sigma/d\Omega$ measurement.

{\it Photon beam asymmetry}--For the measurement of $\Sigma$'s, the event selection was tightened by additionally requiring that a charged track reconstructed at DC should pass through the region within a radius of $640$~$\mathrm{mm}$ from the beam axis on the sensitive area. This condition was applied to avoid any possible bias of proton detection in the azimuthal direction arising from the hexagonal shape of DC. Signal counts were obtained in 7 energy and 5 polar angle bins for $1.896 < W < 2.316$~$\mathrm{GeV}$ and $-1 < \cos \theta^{c.m.}_{\eta^\prime} < 1$, respectively. Here, each kinematic-bin sample to extract a signal count by the template fit explained earlier was divided into two sub-samples, where the azimuthal angle of a reaction plane was parallel and perpendicular to the photon beam polarization vector. Further details of the method to evaluate $\Sigma$'s from the signal counts in these sub-samples are described in Ref.~\cite{f0egg}. In the present analysis, two sets of data were collected by directing the linear polarization vector vertically and horizontally. The $\Sigma$'s were measured by adding these two data sets after redefining the polarization vector directions as a reference angle. Because the statistical consistency of the results from the $2\gamma$ and $6\gamma$ modes was confirmed, they were averaged with weights in the same way as the $d\sigma/d\Omega$ measurement. Any acceptance correction was not needed for the evaluation of $\Sigma$'s.

The systematic uncertainties of $\Sigma$'s were estimated as the absolute deviation from the measured value in each kinematic bin, caused by the following sources. Small deterioration of the azimuthal symmetry in the detector setup should be cancelled by adding the vertical and horizontal polarization data mentioned above, but the difference of $\Sigma$'s between the measurements using the individual data was conservatively evaluated as a systematic uncertainty ($0.0030$--$0.0515$). The ambiguity of photon beam polarization directions \cite{pi0egg} was also turned out to cause small variations in $\Sigma$'s ($0.0003$--$0.0118$), whereas the systematic uncertainties due to the ambiguity of polarization degrees were negligible. Furthermore, possible difference of the background shapes in the two sub-samples with the reaction plane parallel and perpendicular to the polarization vector was considered to care the $\Sigma$'s of background processes. The fixed background template shapes were thus replaced to a second-order polynomial function with free parameters to estimate the $\Sigma$ variation for $\eta^\prime$ photoproduction signals ($0.0003$--$0.0251$). All of the above uncertainties were estimated by combining neighboring kinematic bins to increase the statistics. The uncertainties obtained in the $2\gamma$ and $6\gamma$ modes were merged for each source using the weights based on the statistical uncertainties of $\Sigma$'s. The total systematic uncertainties were finally calculated to be $0.0134$--$0.0517$ by taking a quadratic sum of the listed uncertainties.

\begin{figure*}[tp]
 \centering
 \includegraphics[width=15cm]{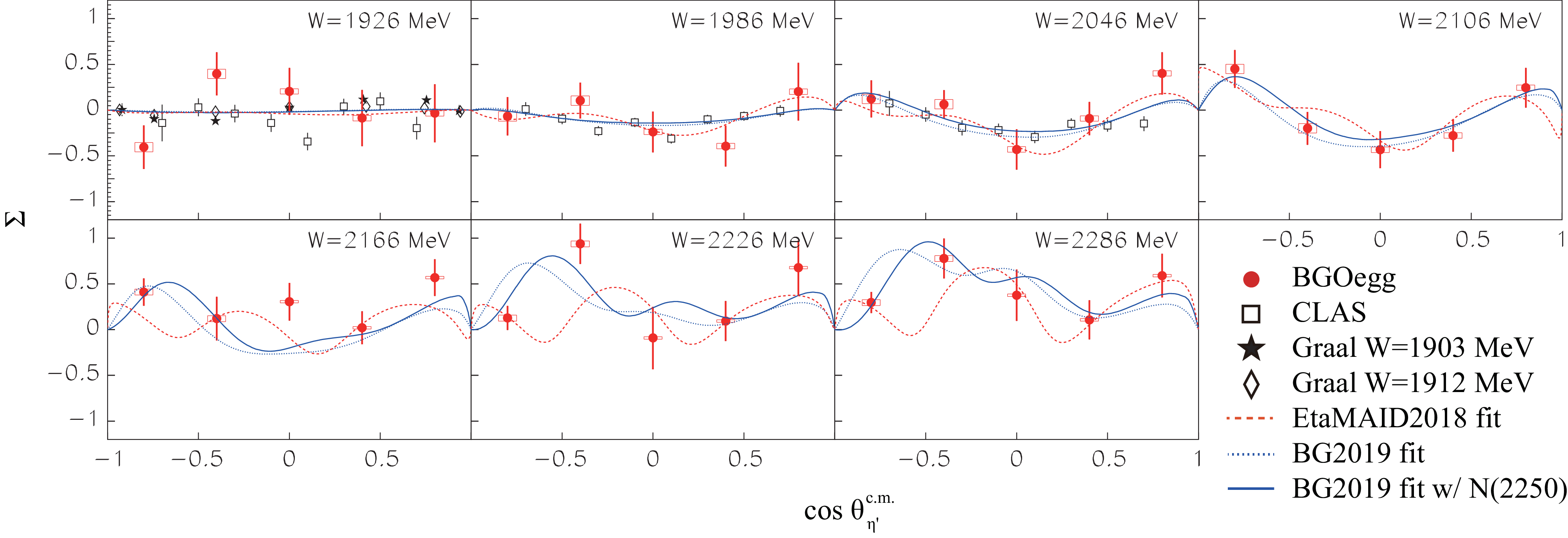}
 \caption{Photon beam asymmetries $\Sigma$ of the reaction $\gamma p \to \eta^\prime p$, plotted for the present analysis (BGOegg) and the existing results indicated in the legend. The notations are the same as those in Fig.~\ref{fig:dcs} except for showing systematic uncertainties by half the vertical lengths of boxes around the data points.}
 \label{fig:pba}
\end{figure*}
Red closed circles in Fig.~\ref{fig:pba} show the $\Sigma$'s measured by the present analysis. Half the lengths of thick vertical bars and open boxes around the data points represent the statistical and systematic uncertainties, respectively. The existing results from the Graal \cite{pbagraal} and CLAS \cite{pbaclas} collaborations are overlaid in low-energy panels using the symbols indicated in the legend. The plotted results are statistically consistent with each other in the overlapped energy regions, whereas the present analysis provides new information at the unexplored energies above $W \approx 2.1$~$\mathrm{GeV}$. In Fig.~\ref{fig:pba}, the measured $\Sigma$'s show relatively small absolute values at lower energies. Because of the statistical reason, the present analysis cannot confirm the angular-dependent behavior of $\Sigma$'s reported by the Graal collaboration. On the other hand, the values in the middle angle range become negative at $W \approx 2.1$~$\mathrm{GeV}$. This behavior is the same as the observation by the CLAS collaboration, which has discussed the evidence for $N(1895)$, $N(1900)$, $N(2100)$, and $N(2120)$ using a Bonn-Gatchina (BG) partial wave analysis (PWA) framework. In the higher energy region, the $\Sigma$'s are positive at most of the polar angles but showing the angular dependence where a backward bin gives higher values.

{\it Partial wave analyses}--Finally, the present results of $d\sigma/d\Omega$'s and $\Sigma$'s were fitted by two PWA models: EtaMAID2018 \cite{em2018} and BG2019 \cite{bg2019}, where the accumulated experimental data were simultaneously input. In these fits, the weights for the present data were increased to inspect their physics implication with the reduction of the relative influence from the other data: 25 times for the $d\sigma/d\Omega$ and $\Sigma$ data in the EtaMAID2018 fit, and 30 times for the $\Sigma$ data in the BG2019 fit. The fitted results are shown by red dashed and blue dotted lines for EtaMAID2018 and BG2019, respectively, in Figs.~\ref{fig:dcs} and \ref{fig:pba}. Both fits suggested that the PWA curves for $d\sigma/d\Omega$'s were insensitive to the inclusion of the present data, more or less reproducing them. In contrast, the PWA curves for $\Sigma$'s were largely changed by the fit to the present data, particularly at backward angles and high energies. The reduced $\chi^2$'s for the present data were improved to $1.7$ and $1.8$ in the new fits of EtaMAID2018 and BG2019, respectively. 

In the EtaMAID2018 fit, masses, widths, coupling constants to $\eta^\prime N$, and proton helicity amplitudes $A_{1/2}$ were made free for the contributions from $N(1895)$, $N(1900)$, $N(1990)$, $N(2000)$, $N(2060)$, $N(2100)$, $N(2120)$, $N(2190)$, and $N(2250)$, which sit in the energy range around the $\eta^\prime$ production threshold or higher. Most of these parameters were not significantly changed by the fit, but only the $\eta^\prime N$ coupling constant of $N(2250)$ possessing $J^P=\frac{9}{2}^-$ increased from $0.085$ to $0.598$. On the other hand, the BG2019 fit took into account the same $N^*$ contributions as EtaMAID2018 but except for $N(2250)$. In this case, the fit to the present data influenced $t$-channel exchange parameters but did not vary the resonance parameters. To further investigate the observed difference of resonance parameter changes, the BG2019 model was modified to include a $J^P=\frac{9}{2}^-$ resonance corresponding to $N(2250)$ and perform a refit in the same way mentioned above. The modified fit significantly improved the reduced $\chi^2$ from $1.8$ to $0.9$ for the present $\Sigma$ data. The results of this BG2019 fit with $N(2250)$ are indicated by a blue solid line in Figs.~\ref{fig:dcs} and \ref{fig:pba}. Here, note that there are differences between the EtaMAID2018 and BG2019 fit results even by including the same resonances because of different minima in their solutions arising from the difference of modeling procedures and the unavailability of a so-called ``complete data'' set \cite{baryspec}. In any case, the above results may imply the possible importance of the $N(2250)$ contribution in $\eta^\prime$ photoproduction, and further confirmation with high statistics data is desired. The necessity of extra resonances was not found in both fits with the current level of statistical uncertainties in the $\Sigma$ data.

{\it Summary}--The comprehensive measurements of $d\sigma/d\Omega$, $\sigma_{tot}$, and $\Sigma$ for $\eta^\prime$ photoproduction off the proton were performed by using a linearly polarized photon beam up to $E_\gamma = 2.4$~$\mathrm{GeV}$ and the detector setup mainly using the BGOegg calorimeter at the SPring-8 LEPS2 beamline. The signal statistics were doubled by analyzing the two $\eta^\prime$ decay modes into $\gamma \gamma$ and $\pi^0 \pi^0 \eta \to \gamma \gamma \gamma \gamma \gamma \gamma$ over all the polar angle range. New constraints for the $N^*$ search were obtained in $\Sigma$'s at unexplored energies and $d\sigma/d\Omega$'s at extremely backward angles. The PWA fits to the present results with increased weights may suggest more contribution from the high-spin resonance $N(2250)$. Near-future analyses providing higher precisions can be expected from the unanalyzed data with similar statistics and the high-statistics data being collected with the calorimeter setup covering a larger solid angle \cite{had2023}.

\section*{Acknowledgments}
The experiment was performed at the BL31LEP of SPring-8 with the approval of the Japan Synchrotron Radiation Institute (JASRI) as a contract beamline (Proposal No.~BL31LEP/6101). The authors gratefully acknowledge to the support of the staff at SPring-8 for providing excellent experimental conditions. This research was supported in part by the Ministry of Education, Culture, Sports, Science and Technology of Japan, JSPS KAKENHI Grant Nos.~19002003, 24244022, 21H04986, the Ministry of Science and Technology of Taiwan, and the RNF grant 24-22-00322.


\end{document}